# Prediction of High-Temperature Half Quantum Anomalous Hall Effect in a Semi-magnetic Topological Insulator of MnBi$_2$Te$_4$/Sb$_2$Te$_3$

M. U. Muzaffar,[1,2] Kai-Zhi Bai,[3] Wei Qin,[4] Guohua Cao,[1,2] Yutong Yang,[1,2] Shunhong Zhang,[1,2] Ping Cui,[1,2,*] Shun-Qing Shen,[3,*] and Zhenyu Zhang[1,2,*]

[1]*International Center for Quantum Design of Functional Materials (ICQD), Hefei National Research Center for Physical Sciences at the Microscale, University of Science and Technology of China, Hefei, Anhui 230026, China*

[2]*Hefei National Laboratory, University of Science and Technology of China, Hefei, Anhui 230088, China*

[3]*Department of Physics, The University of Hong Kong, China*

[4]*Department of Physics, University of Science and Technology of China, Hefei, Anhui 230026, China*

*E-mail: cuipg@ustc.edu.cn; sshen@hku.hk; zhangzy@ustc.edu.cn

## ABSTRACT

The classic Thouless-Kohmoto-Nightingale-Nijs theorem dictates that a single electron band of a lattice can only harbor an integer quantum Hall conductance as a multiple of $e^2/h$, while recent studies have pointed to the emergence of half quantum anomalous Hall (HQAH) effect, though the underlying microscopic mechanisms remain controversial. Here we propose an ideal platform of MnBi$_2$Te$_4$/Sb$_2$Te$_3$ that allows not only to realize the HQAH effect at much higher temperatures, but also to critically assess the different contributions of the gapped and gapless Dirac bands. We first show that the top surface bands of the Sb$_2$Te$_3$ film become gapped, while the bottom surface bands remain gapless due to proximity coupling with the MnBi$_2$Te$_4$ overlayer. Next we show that such a semi-magnetic topological insulator harbors the HQAH effect at ~20 K, with Cr doping enhancing it to as high as 67 K, driven by large magnetic anisotropy and strong magnetic coupling constants that raise the Curie temperature. Our detailed Berry curvature analysis further helps to reveal that, whereas the gapped surface bands can contribute to the Hall conductance when the chemical potential is tuned to overlap with the bands, these bands have no net contribution when the chemical potential is in the gapped region, leaving the gapless bands to be the sole contributor to the HQAH conductance. Counterintuitively, the part of the gapless bands within the gapped region of the top surface bands have no net contribution, thereby ensuring the plateau nature of the Hall conductance.

Topological materials are characterized by their nontrivial band structures in bulk form with robust surface states[1-3]. One prime example of such quantum states of matter is the quantum anomalous Hall (QAH) insulator [4], characterized by the integer quantum Hall conductance as multiples of $e^2/h$ (where $e$ and $h$ denote the electron charge and Planck's constant, respectively) [5]. Unlike the conventional quantum Hall effect [6], the QAH effect does not require an external magnetic field, offering a fertile ground for exploring exotic quantum phenomena with potential applications in low-power or dissipationless electronic devices. Since its initial conceptualization [4], substantial efforts have been made to identify enabling platforms for realizing the QAH effect [7-16]. One approach is to induce magnetism in topological insulators (TIs) through magnetic doping [7], which has led to the first observed QAH effect in Cr/V-doped $(Bi,Sb)_2Te_3$ films [11-14]. Recently, such nontrivial states have also been observed in an intrinsic magnetic TI of $MnBi_2Te_4$ at elevated temperatures [17,18], offering exciting prospects for future practical applications.

Aside from the QAH effect with Hall conductivity of integer multiples of $e^2/h$, strongly correlated fermionic systems may also harbor fractional quantum anomalous Hall (FQAH) states at zero magnetic field [19-22], analogue to the fractional quantum Hall effects observed in partially filled Landau levels established at high magnetic fields [23,24]. These intriguing phenomena have been rationalized within the framework of composite fermions [25,26], with the distinction that the harboring systems are uniquely insulating. Very recently, a special type of the FQAH effect has been observed in Cr-doped $(Bi,Sb)_2Te_3$ with a half quantum anomalous Hall (HQAH) conductance of $e^2/2h$ at ~1 K [27,28], but the underlying physics remains controversial. Clearly, given the semi-metallic nature of the system, the composite fermionic picture is not applicable here. One interpretation is to attribute the HQAH conductance to the gapped surface states of the TI [27,28]. A competing mechanism is that, since a gapped band on a lattice can only harbor integer quantum Hall conductance of 0 or 1 $e^2/h$ [5], the HQAH conductance is due to the existence of the single gapless Dirac cone of the surface band [29-32].

In this work, we use first-principles calculations to study an ideal platform of $MnBi_2Te_4/Sb_2Te_3$ that allows not only to realize the HQAH effect at much higher temperatures, but also to critically assess the different contributions of the gapped and gapless Dirac bands. We first show that the top surface bands of the $Sb_2Te_3$ film become gapped, while the bottom surface bands remain gapless due to proximity coupling with the $MnBi_2Te_4$ overlayer. Next we show that such a semi-magnetic TI harbors the HQAH effect at ~20 K, with Cr doping enhancing it to as high as 67 K, driven by large magnetic anisotropy and strong magnetic coupling constants that raise the Curie temperature ($T_c$). Our detailed Berry curvature analysis further helps to reveal that, whereas the gapped surface bands can contribute to the Hall conductance when the chemical potential is tuned to overlap with the bands, these bands have no net contribution when the chemical potential is in the gapped region, leaving the gapless bands to be the sole contributor to the HQAH conductance. Counterintuitively, the part of the gapless bands within the gapped region of the top surface bands have no net contribution, thereby ensuring the plateau nature of the Hall conductance. These findings not only offer much-needed clarifying insights into HQAH conductance but also propose a general framework for identifying new candidate materials exhibiting the HQAH effect.

First-principles calculations based on density functional theory (DFT) were performed using the projector-augmented wave method [33], implemented in the VASP code [34]. The Perdew-Burke-Ernzerhof (PBE) exchange-correlation functional with the generalized gradient approximation (GGA)+$U$ method was used to treat the localized $d$ orbitals [35]. To compute the anomalous Hall conductivity, the maximally localized Wannier functions were constructed using the WANNIER90 code [36] in conjunction with the WannierTools package [37]. More computational details, along with the relevant references [33-41], are provided in the Supplemental Material [42].

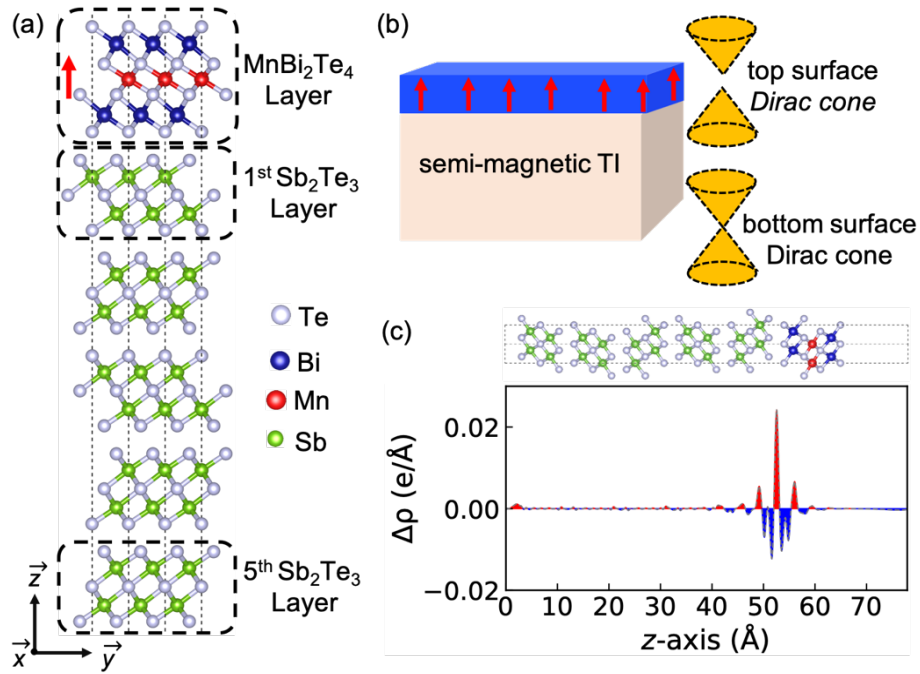

**FIG. 1.** (a) Atomic structure of the MnBi$_2$Te$_4$/Sb$_2$Te$_3$ heterostructure, with the red arrow indicating the Mn 3$d$ moments pointing out of the surface plane. (b) Schematic representation

of a semi-magnetic TI with a gapped Dirac cone on the top surface and a gapless Dirac cone on the bottom surface. (c) The *xy*-plane-averaged charge density difference, with the atomic structure shown in the top panel. The red and blue regions represent charge accumulation and depletion, respectively.

The tetradymite-type MnBi$_2$Te$_4$ compound crystallizes in a rhombohedral layered structure, similar to Bi$_2$Te$_3$, with the space group $R\bar{3}m$ (No. 166) [43]. This layered compound can be viewed as a structure consisting of a Bi$_2$Te$_3$ layer intercalated with an additional Mn-Te layer. Within a single septuple layer (SL) (Te-Bi-Te-Mn-Te-Bi-Te), the Mn atoms exhibit ferromagnetic order, while adjacent SLs exhibit antiferromagnetic order [43-47]. Due to weak van der Waals interactions, MnBi$_2$Te$_4$ can be mechanically exfoliated into septuple layers, and the exfoliated thin films exhibit *n*-type semiconducting behavior [17]. To construct the heterostructure, we choose Sb$_2$Te$_3$ from among various 3D TIs due to its intrinsic *p*-type characteristics [48]. Considering the intrinsic *p*- and *n*-type conducting characteristics of the two materials, the MnBi$_2$Te$_4$/Sb$_2$Te$_3$ heterostructure should serve as an ideal platform for natural charge compensation, a key aspect for achieving quantized transport phenomena at higher temperatures [16]. Figure 1(a) presents the optimized crystal structure of the MnBi$_2$Te$_4$/Sb$_2$Te$_3$ heterostructure, with an interlayer distance of 2.96 Å, which is comparable to distance between adjacent SLs in bulk MnBi$_2$Te$_4$ (2.76 Å), suggesting a vdw-type interaction. To evaluate the energetic viability of the proposed heterostructure, we calculate the binding energy, defined as $E_b = E_{total} - E_{ST} - E_{MBT}$, with $E_{total}$, $E_{ST}$, and $E_{MBT}$ being the total energies of the heterostructure, the freestanding five quintuple layers (QLs) of Sb$_2$Te$_3$, and the freestanding MnBi$_2$Te$_4$ SL, respectively. The calculated binding energy is -26.32 meV/Å$^2$, in the same order of magnitude as that of other vdW heterostructures, such as InSe/GeSe(S) [49]. Charge transfer analysis shows that depositing MnBi$_2$Te$_4$ on Sb$_2$Te$_3$ induces interfacial charge accumulation, as shown in Fig. 1(c), where the charge density difference (Δρ) is the plane-averaged difference along the z-axis between the heterostructure and the sum of the isolated films. Moreover, our *ab initio* molecular dynamics simulations at 200 K and up to 5 ps demonstrate that the system is thermodynamically stable as well, as shown in Fig. S1. Thus, we can anticipate that the proposed heterostructure is both energetically and thermodynamically stable and can be synthesized experimentally.

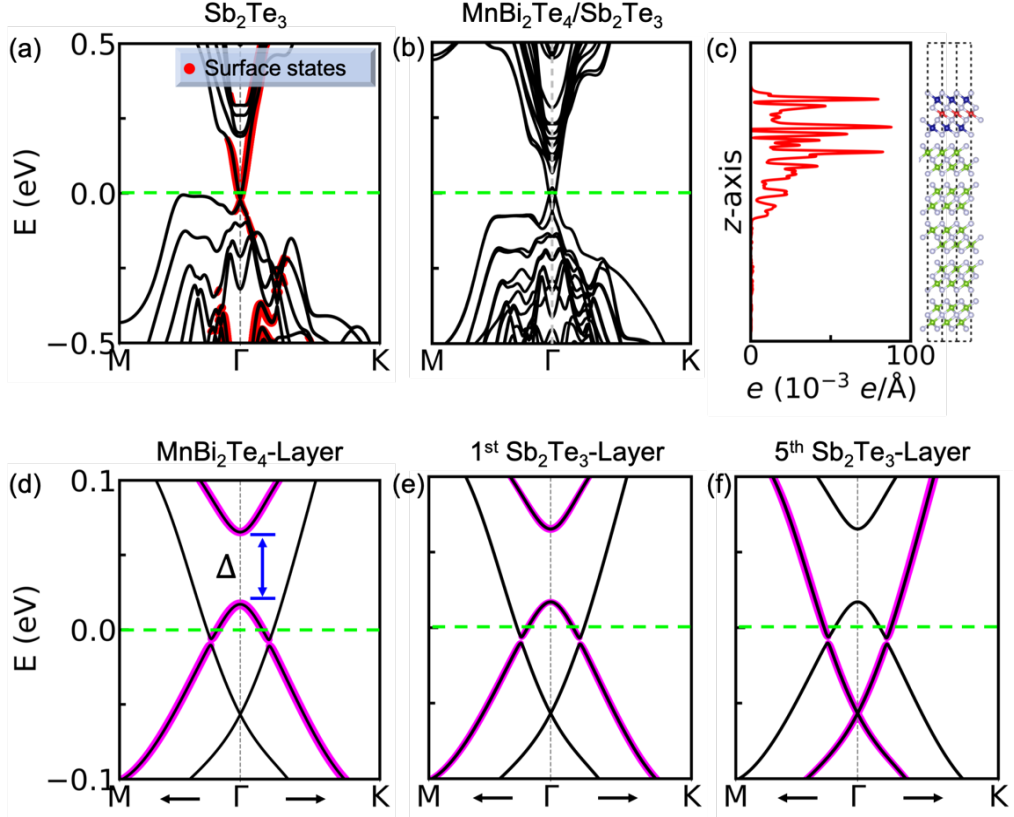

**FIG. 2.** Band structures of (a) 5-QL $Sb_2Te_3$ with the red circles denoting the projection onto the topmost QL and (b) $MnBi_2Te_4$/5-QL $Sb_2Te_3$ heterostructure. (c) Electron density distributions of the gapped topological surface states at the $\Gamma$ point for the $MnBi_2Te_4$/$Sb_2Te_3$ heterostructure, with the corresponding crystal structure shown in the right panel. (d-f) Zoomed-in views of the layer-resolved band structures of the $MnBi_2Te_4$/$Sb_2Te_3$ heterostructure, with the top-surface Dirac-cone gap shown as $\Delta$. The contributions of the selected layers, as highlighted in Fig. 1(a), are characterized by the size of the magenta circles.

Figure 2(a) illustrates the band structure of 5-QL $Sb_2Te_3$ along the high-symmetry lines of the 2D Brillouin zone, computed with spin-orbit coupling. By projecting the wave functions of the topmost or bottommost QL onto the band structure [50], the surface states can be identified. It is evident that the topologically nontrivial surface states cross the Fermi level, suggesting that the chosen film thickness is sufficient to decouple the top and bottom surface states. The critical thickness for this decoupling is $d \geq 4$ QLs [51], which is essential for realizing the HQAH effect, as confirmed in Figs. S3 and S4. Upon forming the $MnBi_2Te_4$/$Sb_2Te_3$ heterostructure, considerable changes can be seen in the band structure, as shown in Figs. 2(b) and 2(d)-2(f). For instance, obvious band splittings below the Fermi level around the $\Gamma$ point are observed, which can be mainly attributed to the structural asymmetry of the MnBi$_2$Te$_4$/Sb$_2$Te$_3$ heterostructure. More importantly, the exchange interaction between the surface electrons and magnetism induces an energy gap in the top-surface Dirac cone while keeping the bottom-surface Dirac cone gapless [see Figs. 2 (d)-2(f)]. Here, an analysis of the projected layer-resolved band structures shows that the low-energy states of the top-surface gapped Dirac cone primarily arise from hybridizations between the MnBi$_2$Te$_4$ layer and the 1$^{st}$ Sb$_2$Te$_3$ layer, while the 5$^{th}$ Sb$_2$Te$_3$ layer still contributes to those of the bottom-surface gapless Dirac cone. Such a system can be referred to as a semi-magnetic TI, and the magnetic gap ($\Delta$) of the top-surface Dirac cone is calculated to be 48.2 meV, larger than the thermal energy at room temperature. Since it is well-established that PBE approximation underestimates the bandgap, we employ the SCAN functional at the meta-GGA level to accurately predict the magnetic gap [52]. As shown in Fig. S6, the SCAN functional produces a band structure similar to that of PBE but enhances the magnetic gap of the top-surface Dirac cone from 48.2 meV to 83 meV, thereby validating our PBE results. Such a large magnetic surface gap is attributed to the strong interaction between the surface states of the TI and the magnetism (see Fig. 2(c)). Moreover, it is noted that there is a fake gap below the Fermi level, possibly due to the quantum confinement effect [53], which decreases with thicker Sb$_2$Te$_3$ layers; however, due to computational constraints, we restrict the analysis to up to 5 QLs.

From a realistic perspective, it is widely recognized that MnBi$_2$Te$_4$ exhibits intrinsic antisite defects, typically manifesting as Mn-Bi intermixing during the crystal growth [54,55]. Such unwanted crystal defects may influence the energy gap in the top-surface bands of TI. To ensure the robustness of the magnetic gap against Mn-Bi intermixing, we construct a $\sqrt{3}\times\sqrt{3}$ supercell of the MnBi$_2$Te$_4$/Sb$_2$Te$_3$ heterostructure, where one Bi atom is replaced with Mn and one Mn atom is replaced with Bi (see Fig. S7). Clearly it can be seen that the overall shapes of the bands with and without Mn-Bi intermixing remain same. Strikingly, even with such a high concentration of Mn-Bi intermixing, a magnetic gap of 7 meV can still persist in the top-surface Dirac cone. Moreover, such a magnetic gap is tuneable upon in-plane strain, as shown in Fig. S5(b).

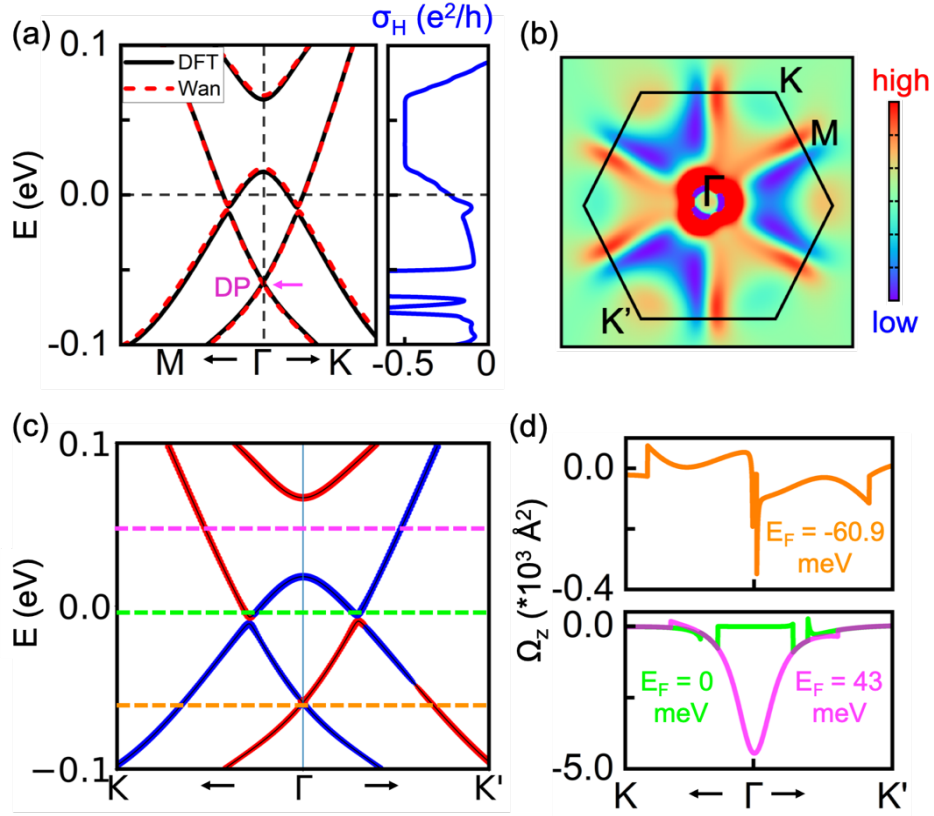

**FIG 3.** (a) Electronic band structure (left panel) and Hall conductivity as a function of energy (right panel) for the MnBi$_2$Te$_4$/Sb$_2$Te$_3$. The magenta arrow indicates the gapless Dirac point. (b) The corresponding Berry curvature distribution in the 2D Brillouin zone. (c) Band-resolved and (d) Fermi-energy-dependent Berry curvatures along the K→Γ→K' path. In (c), the red and blue circles represent positive and negative values of the Berry curvature, respectively, while the horizontal dashed lines correspond to the chosen Fermi level positions (E$_F$) for (d).

Now, we turn our attention to the central issue of this study, which is the quantized Hall conductivity at zero temperature, calculated by means of the following formula [56],

$$\sigma_H = -\frac{e^2}{h}\int_{BZ}\frac{dk_x dk_y}{2\pi}\Omega_z(k_x,k_y). \quad (1)$$

Here the Berry curvature can be defined as

$$\Omega_z(k_x,k_y) = -2\operatorname{Im}\sum_n^{occupied}\sum_{n'}^{empty}\frac{\langle\Psi_{nk}|\hbar v_x|\Psi_{n'k}\rangle\langle\Psi_{n'k}|\hbar v_y|\Psi_{nk}\rangle}{(\varepsilon_{nk}-\varepsilon_{n'k})^2}, \quad (2)$$

where $\Psi_{nk}$ is the Bloch wave function with eigenvalues $\varepsilon_{nk}$ and $v_x(v_y)$ is the velocity operator along the $x$ ($y$) direction. The calculated Hall conductivity for the proposed heterostructure is plotted in Fig. 3(a). Clearly, in the absence of an external magnetic field, the HQAH effect can be realized in MnBi$_2$Te$_4$/Sb$_2$Te$_3$ heterostructure with a Hall conductivity of –0.5 $e^2/h$. Interestingly, flipping the magnetization of the system reverses the sign of the Berry curvature at the Γ point [Fig. 3(a,b) and Fig. S8], enabling a switchable Hall conductivity of +0.5 $e^2/h$. It is noted that there is an anomaly in the Hall conductance around the Dirac cone of the gapless surface state, which is caused by a fake gap, similar to previous discussion. Here, it is worth noting that, compared to the oscillatory Hall conductivity around $e^2/2h$ in the asymmetric MnBi$_2$Te$_4$/(Bi$_2$Te$_3$)$_n$/MnBi$_2$Te$_4$ configuration [57], the switchable and quantized Hall conductivity (±0.5 $e^2/h$) over a broad energy range in MnBi$_2$Te$_4$/Sb$_2$Te$_3$ semi-magnetic TI offers significant practical advantages for device applications.

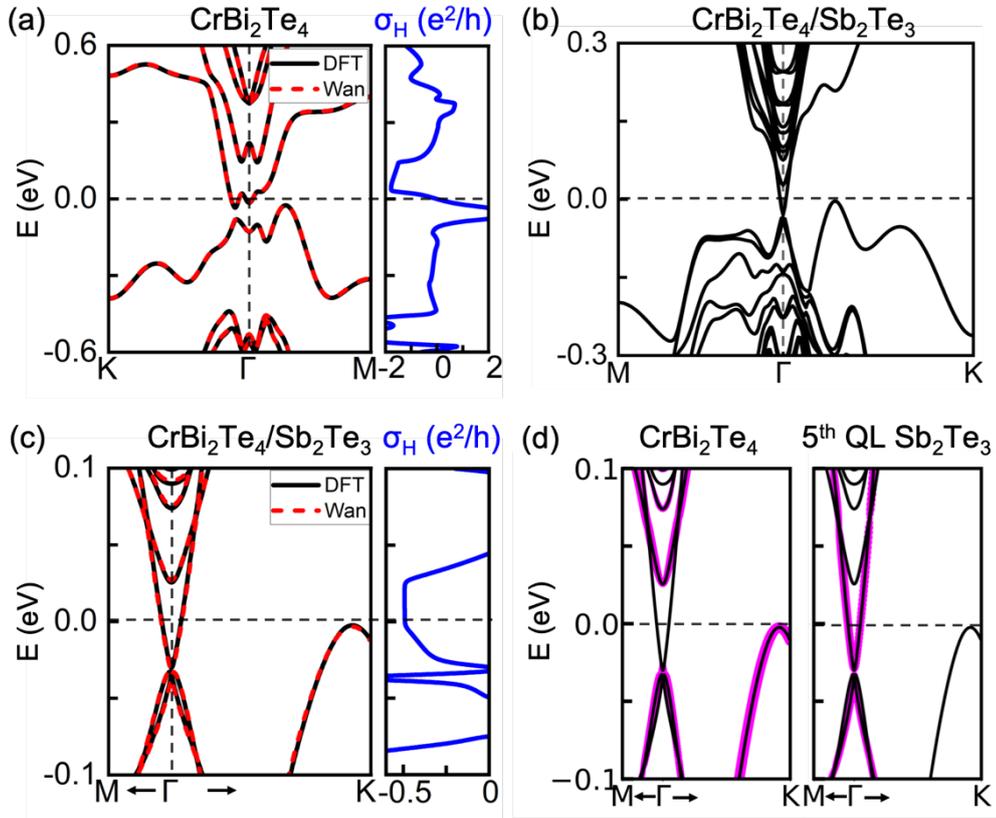

**FIG. 4.** Electronic band structures (left panels) and Hall conductivities as functions of energy (right panels) for (a) freestanding CrBi$_2$Te$_4$, and (c) a zoomed-in view of panel (b) of the CrBi$_2$Te$_4$/Sb$_2$Te$_3$ heterostructure. (d) Layer-resolved band structures of the CrBi$_2$Te$_4$/Sb$_2$Te$_3$ heterostructure, with the contributions of selected layers indicated by the size of the magenta circles.

To realize HQAH conductivity in our proposed heterostructure, the Fermi level needs to be tuned via chemical doping [11] or by electrolytic/ionic liquid gating techniques [58,59]. To achieve this, we demonstrate that stacking $CrBi_2Te_4$ on top of $Sb_2Te_3$ can shift the gapped Dirac cone towards lower energy, thereby enabling HQAH conductivity at the Fermi level (see Fig. 4(c)). In stark contrast, an isolated $CrBi_2Te_4$ exhibits metallic behavior (see Fig. 4(a)), which may explain why it has not been as actively explored as $MnBi_2Te_4$, particularly in the context of quantized transport phenomena. Here, it is noted that the proposed $CrBi_2Te_4/Sb_2Te_3$ heterostructure is both energetically ($E_b = -26.46$ meV/Å²) and thermodynamically stable (see Fig. S2). Therefore, the proposed approach for realizing the HQAH effect is generic and applicable to other related 2D semimagnetic TIs, regardless of whether the ferromagnetic layer in such systems is metallic or semiconducting—provided that its magnetization is strong enough to induce a magnetic gap in one surface Dirac cone.

Having confirmed the HQAH effect in $MnBi_2Te_4/Sb_2Te_3$, our next objective is to investigate the detailed contributions of the gapped and gapless Dirac surface bands to the HQAH conductivity. In doing so, we calculate the Berry curvature distribution because the Hall conductivity, as per Eq. (1), is contributed by the $\Omega_z(k)$ for the occupied states at each $k$-point in the first Brillouin zone. Before delving into Berry curvature analysis, note that the band of the surface Dirac cone typically splits into two parts within the Brillouin zone: the low-energy part exhibits linear energy dispersion near its center, describing massless Dirac fermions, while the high-energy part, farther from the center, displays non-linear behavior characteristic of massive particles. We calculate the band-resolved and Fermi-energy-dependent Berry curvatures along the K→Γ→K' path in the low-energy region. As shown in Fig. 3(c), the Berry curvatures of the gapless bands exhibit opposite signs near the Γ center along Γ→K and Γ→K', potentially resulting in no overall contribution to the Hall conductance. In contrast, the Berry curvatures of the gapped bands are unable to cancel each other out, leading to a non-zero contribution when the chemical potential is tuned to overlap with the bands. As illustrated in Fig. 3(d) and Fig. S10, adjusting the Fermi level within the gapped Dirac cone region can lead to a significantly large Berry curvature value around the Γ point, which is invariant upon shifting the chemical potential inside the magnetic gap. It is worth mentioning that the Fermi energy-dependent Berry curvature analysis is experimentally feasible. For example, this can be achieved through electrical gating technique, which generally induces rigid shifts of the Fermi level without significantly altering the band dispersion. As a consequence, the dominant physical picture remains unchanged [60]. Moreover, our spin-dependent band structure calculations show that the spin orientation remains unchanged within the specific energy window considered (see Fig. S11). Based on these calculations, we can conclude that, in the low-energy region, the surface states of the gapless Dirac cone do not contribute to the Hall conductance, while magnetically gapped states of the gapped Dirac cone have a net contribution.

To further clarify the origin of the HQAH conductance in $MnBi_2Te_4/Sb_2Te_3$, the high-energy region needs to be taken into account. Given the fact that the classic Thouless-Kohmoto-Nightingale-Nijs theorem dictates that a single electron band of a lattice can only harbor an integer quantum Hall conductance of 0 or 1 $e^2/h$ [5], the gapped surface bands contribute to approximately HQAH conductivity at low energy near the center of the Brillouin zone, which should be counterbalanced by another nearly HQAH conductivity of the bands at high energy in the Brillouin zone. Combined with our previous understanding via the tight-binding lattice models, since the gapped bands contribute to a Hall conductance at low energy, this conductivity should be compensated by a conductance from the states at high energy [30,31]. Therefore, these gapped bands have no net contribution when the chemical potential is in the gapped region, leaving the gapless bands to be the sole contributor to the HQAH conductance. Here we would like to emphasize that, counterintuitively, the gapless bands generate HQAH conductivity with high-energy symmetry-broken states, but have no net contribution from parity symmetric surface states within the gapped region of the top surface states, thereby ensuring the plateau nature of the Hall conductivity.

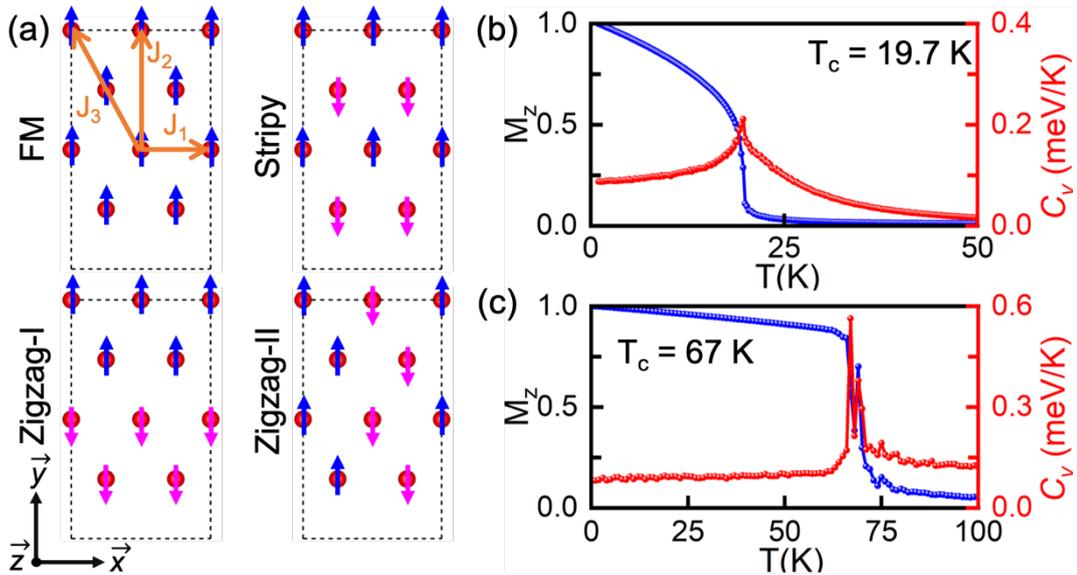

**FIG 5.** (a) Four different spin configurations of a $MnBi_2Te_4$ ($CrBi_2Te_4$) SL on top of 3-QL $Sb_2Te_3$. For clarity, only the Mn (Cr) atoms are shown here, with the opposite spin directions

indicated by the magenta and blue arrows, while the orange arrows denote the exchange interactions. Magnetization ($M_z$) and specific heat capacity ($C_v$) as functions of temperature ($T$) for the (b) MnBi$_2$Te$_4$/Sb$_2$Te$_3$ and (c) CrBi$_2$Te$_4$/Sb$_2$Te$_3$ heterostructures.

Given that MnBi$_2$Te$_4$/Sb$_2$Te$_3$ and CrBi$_2$Te$_4$/Sb$_2$Te$_3$ semi-magnetic TIs exhibit the HQAH effect, the next step is to estimate the Hall temperature based on the nontrivial band gap and the $T_c$ of magnetic ordering. To rigorously calculate the $T_c$, we have performed the Monte Carlo simulations. The Heisenberg Hamiltonian can be written as:

$$H = -\frac{1}{2}\sum_{ij} J_{ij} \vec{S}_i \cdot \vec{S}_j - D\sum_i S_i^2, \qquad (3)$$

where $J_{ij}$ is the magnetic coupling constant between sites $i$ and $j$, $D$ is the single-ion anisotropy parameter, and $S$ is the spin operator at site $i$. Positive (negative) values of $J$ correspond to ferromagnetic (antiferromagnetic) interactions. A (2 × 2√3) supercell size is used to calculate the total energies of four competing magnetic configurations, corresponding to one ferromagnetic and three antiferromagnetic states, as shown in Fig 5(a). We mainly consider the nearest-, second-nearest-, and third-nearest-neighbour exchange interactions, denoted as $J_1$, $J_2$, and $J_3$, respectively. By examining the total energies of the four spin configurations, the magnetic coupling constants can be extracted from the following equations:

$$\begin{aligned} E_{FM} &= E_0 - 24J_1|\vec{S}|^2 - 24J_2|\vec{S}|^2 - 24J_3|\vec{S}|^2, \\ E_{stripy} &= E_0 + 8J_1|\vec{S}|^2 + 8J_2|\vec{S}|^2 - 24J_3|\vec{S}|^2, \\ E_{zigzag1/zigzag2} &= E_0 \mp 8J_1|\vec{S}|^2 \pm 8J_2|\vec{S}|^2 + 8J_3|\vec{S}|^2. \end{aligned} \qquad (4)$$

The calculated exchange constants $J_1$, $J_2$, and $J_3$ for MnBi$_2$Te$_4$/Sb$_2$Te$_3$ are 1.46, –0.012, and 0.028 meV, respectively, suggesting that $J_1$ play a dominant role in establishing the ferromagnetic ground state [61,62]. Interestingly, stacking CrBi$_2$Te$_4$ on top of Sb$_2$Te$_3$ significantly enhances the exchange constants ($J_1$ = 11.74 meV, $J_2$ = 6.58 meV and $J_3$ = –9.74 meV), suggesting that the competition between $J_1$ and $J_3$ would play an important role in determining the $T_c$ of the CrBi$_2$Te$_4$/Sb$_2$Te$_3$ heterostructure. We further calculate the magnetic anisotropy energy ($E_{MAE}$), defined as the total energy difference between the in-plane and out-of-plane magnetic moment configurations ($E_{MAE} = E_{in} - E_{out}$), meanwhile taking the spin-orbit coupling effect into account. Compared to an isolated MnBi$_2$Te$_4$, the MnBi$_2$Te$_4$/Sb$_2$Te$_3$ heterostructure exhibits an increased $E_{MAE}$ from 0.11 to 0.14 meV per Mn atom while maintaining a preference for out-of-plane magnetization. In the case of the CrBi$_2$Te$_4$/Sb$_2$Te$_3$ heterostructure, the $E_{MAE}$ is significantly higher, reaching a value of 12.64 meV per Cr atom. Using these DFT-derived $E_{MAE}$ and $J$ parameters, the $T_c$ can be estimated for the studied heterostructures through Monte Carlo simulations, as demonstrated in Fig. 5(b,c). It can be seen that the magnetization ($M_z$) decreases with the increase in temperature and abruptly drops to zero at 19.7 K, accompanied by a peak in the specific heat capacity ($C_v$). The estimated $T_c$ for the MnBi$_2$Te$_4$/Sb$_2$Te$_3$ heterostructure is about 19.7 K, comparable to or even higher than that of an isolated MnBi$_2$Te$_4$ [61,62], while the CrBi$_2$Te$_4$/Sb$_2$Te$_3$ heterostructure exhibits a strikingly higher $T_c$ of 67 K. Based on these analyses, we may safely conclude that the large magnetic surface gap and high $T_c$ of the proposed CrBi$_2$Te$_4$/Sb$_2$Te$_3$ heterostructure allow for the manifestation of the HQAH effect at temperatures as high as 67 K, which is notably higher than what has been observed in Cr-doped (Bi,Sb)$_2$Te$_3$ films (~1 K) [27].

In short, following the structural and compositional compatibility between MnBi$_2$Te$_4$ and Sb$_2$Te$_3$, we have identified an ideal platform of the semi-magnetic TI, the MnBi$_2$Te$_4$/Sb$_2$Te$_3$ heterostructure, which allows not only to realize the HQAH effect at much higher temperatures, but also to critically assess the different contributions of the gapped and gapless Dirac surface bands. The proposed approach is conceptually general and sheds new light in experimental realization of the HQAH conductivity at high temperatures.

This work was supported by the National Natural Science Foundation of China (Grant Nos., 12374458, 12488101, 11974323, 12004368, 12474134, 12474243 and W2433002), the Innovation Program for Quantum Science and Technology (Grant No. 2021ZD0302800), the Strategic Priority Research Program of Chinese Academy of Sciences (Grant No. XDB0510200), the Anhui Provincial Key Research and Development Project (Grant No. 2023z04020008), the Research Grants Council, University Grants Committee, Hong Kong (Grant Nos. C7012-21G and 17301823), and Quantum Science Center of Guangdong-Hong Kong-Macao Greater Bay Area (Grant No. GDZX2301005).

[1]     M. Z. Hasan and C. L. Kane, Colloquium: Topological insulators, Rev. Mod. Phys. **82**, 3045 (2010).
[2]     Xiao-Liang Qi and Shou-Cheng Zhang, Topological insulators and superconductors, Rev. Mod. Phys. **83**, 1057 (2011).
[3]     F. Duncan M. Haldane, Nobel Lecture: Topological quantum matter, Rev. Mod. Phys. **89**, 040502 (2017).

[4]     F. D. M. Haldane, Model for a quantum Hall effect without Landau levels: Condensed-matter realization of the "parity anomaly", Phys. Rev. Lett. **61**, 2015 (1988).

[5]     D. J. Thouless, M. Kohmoto, M. P. Nightingale, and M. den Nijs, Quantized Hall Conductance in a Two-Dimensional Periodic Potential, Phys. Rev. Lett. **49**, 405 (1982).

[6]     K. v. Klitzing, G. Dorda, and M. Pepper, New method for high-accuracy determination of the fine-structure constant based on quantized Hall resistance, Phys. Rev. Lett. **45**, 494 (1980).

[7]     Rui Yu, Wei Zhang, Hai-Jun Zhang, Shou-Cheng Zhang, Xi Dai, and Zhong Fang, Quantized Anomalous Hall Effect in Magnetic Topological Insulators, Science **329**, 61 (2010).

[8]     Zhenhua Qiao, Shengyuan A. Yang, Wanxiang Feng, Wang-Kong Tse, Jun Ding, Yugui Yao, Jian Wang, and Qian Niu, Quantum anomalous Hall effect in graphene from Rashba and exchange effects, Phys. Rev. B **82**, 161414(R) (2010).

[9]     Wang-Kong Tse, Zhenhua Qiao, Yugui Yao, A. H. MacDonald, and Qian Niu, Quantum anomalous Hall effect in single-layer and bilayer graphene, Phys. Rev. B **83**, 155447 (2011).

[10]    Rui-Lin Chu, Junren Shi, and Shun-Qing Shen, Surface edge state and half-quantized Hall conductance in topological insulators, Phys. Rev. B **84**, 085312 (2011).

[11]    Cui-Zu Chang, Jinsong Zhang, Xiao Feng, Jie Shen, Zuocheng Zhang, Minghua Guo, Kang Li, Yunbo Ou, Pang Wei, Li-Li Wang *et al.*, Experimental observation of the quantum anomalous Hall effect in a magnetic topological insulator, Science **340**, 167 (2013).

[12]    Xufeng Kou, Shih-Ting Guo, Yabin Fan, Lei Pan, Murong Lang, Ying Jiang, Qiming Shao, Tianxiao Nie, Koichi Murata, Jianshi Tang *et al.*, Scale-Invariant Quantum Anomalous Hall Effect in Magnetic Topological Insulators beyond the Two-Dimensional Limit, Phys. Rev. Lett. **113**, 137201 (2014).

[13]    J. G. Checkelsky, R. Yoshimi, A. Tsukazaki, K. S. Takahashi, Y. Kozuka, J. Falson, M. Kawasaki, and Y. Tokura, Trajectory of the anomalous Hall effect towards the quantized state in a ferromagnetic topological insulator, Nature Phys. **10**, 731 (2014).

[14]    Cui-Zu Chang, Weiwei Zhao, Duk Y. Kim, Haijun Zhang, Badih A. Assaf, Don Heiman, Shou-Cheng Zhang, Chaoxing Liu, Moses H. W. Chan, and Jagadeesh S. Moodera, High-precision realization of robust quantum anomalous Hall state in a hard ferromagnetic topological insulator, Nature Mater. **14**, 473 (2015).

[15]    Zhiyong Wang, Chi Tang, Raymond Sachs, Yafis Barlas, and Jing Shi, Proximity-Induced Ferromagnetism in Graphene Revealed by the Anomalous Hall Effect, Phys. Rev. Lett. **114**, 016603 (2015).

[16]    Shifei Qi, Zhenhua Qiao, Xinzhou Deng, Ekin D. Cubuk, Hua Chen, Wenguang Zhu, Efthimios Kaxiras, S. B. Zhang, Xiaohong Xu, and Zhenyu Zhang, High-temperature quantum anomalous Hall effect in n−p codoped topological insulators, Phys. Rev. Lett. **117**, 056804 (2016).

[17]    Yujun Deng, Yijun Yu, Meng Zhu Shi, Zhongxun Guo, Zihan Xu, Jing Wang, Xian Hui Chen, and Yuanbo Zhang, Quantum anomalous Hall effect in intrinsic magnetic topological insulator $MnBi_2Te_4$, Science **367**, 895 (2020).

[18]    Jun Ge, Yanzhao Liu, Jiaheng Li, Hao Li, Tianchuang Luo, Yang Wu, Yong Xu, and Jian Wang, High-Chern-number and high-temperature quantum Hall effect without Landau levels, Natl. Sci. Rev. **7**, 1280 (2020).

[19]    D.N. Sheng, Zheng-Cheng Gu, Kai Sun, and L. Sheng, Fractional quantum Hall effect in the absence of Landau levels, Nat. Commun. **2**, 389 (2011).

[20]    Fan Xu, Zheng Sun, Tongtong Jia, Chang Liu, Cheng Xu, Chushan Li, Yu Gu, Kenji Watanabe, Takashi Taniguchi, Bingbing Tong *et al.*, Observation of Integer and Fractional Quantum Anomalous Hall Effects in Twisted Bilayer $MoTe_2$, Phys. Rev. X **13**, 031037 (2023).

[21]    Heonjoon Park, Jiaqi Cai, Eric Anderson, Yinong Zhang, Jiayi Zhu, Xiaoyu Liu, Chong Wang, William Holtzmann, Chaowei Hu, Zhaoyu Liu *et al.*, Observation of fractionally quantized anomalous Hall effect, Nature **622**, 74 (2023).

[22]    Zhengguang Lu, Tonghang Han, Yuxuan Yao, Aidan P. Reddy, Jixiang Yang, Junseok Seo, Kenji Watanabe, Takashi Taniguchi, Liang Fu, and Long Ju, Fractional quantum anomalous Hall effect in multilayer graphene, Nature **626**, 759 (2024).

[23]    D. C. Tsui, H. L. Stormer, and A. C. Gossard, Two-Dimensional Magnetotransport in the Extreme Quantum Limit, Phys. Rev. Lett. **48**, 1559 (1982).

[24]    Horst L. Stormer, Daniel C. Tsui, and Arthur C. Gossard, The fractional quantum Hall effect, Rev. Mod. Phys. **71**, S298 (1999).

[25]    J. K. Jain, Composite-fermion approach for the fractional quantum Hall effect, Phys. Rev. Lett. **63**, 199 (1989).

[26]    J. K. Jain, Composite Fermion Theory of Exotic Fractional Quantum Hall Effect, Annu. Rev. Condens. Matter Phys. **6**, 39 (2015).

[27]    M. Mogi, Y. Okamura, M. Kawamura, R. Yoshimi, K. Yasuda, A. Tsukazaki, K. S. Takahashi, T. Morimoto, N. Nagaosa, M. Kawasaki *et al.*, Experimental signature of the parity anomaly in a semi-magnetic topological insulator, Nature Phys. **18**, 390 (2022).

[28]    Ryota Watanabe, Ryutaro Yoshimi, Kei S. Takahashi , Atsushi Tsukazaki, Masashi Kawasaki, Minoru Kawamura, and Yoshinori Tokura, Gate-electric-field and magnetic-field

control of versatile topological phases in a semi-magnetic topological insulator, Appl. Phys. Lett. **123**, 183102 (2023).

[29] Bo Fu, Jin-Yu Zou, Zi-Ang Hu, Huan-Wen Wang, and Shun-Qing Shen, Quantum anomalous semimetals, npj Quantum Mater. **7**, 94 (2022).

[30] Jin-Yu Zou, Bo Fu, Huan-Wen Wang, Zi-Ang Hu, and Shun-Qing Shen, Half-quantized Hall effect and power law decay of edge-current distribution, Phys. Rev. B **105**, L201106 (2022).

[31] Jin-Yu Zou, Rui Chen, Bo Fu, Huan-Wen Wang, Zi-Ang Hu, and Shun-Qing Shen, Half-quantized Hall effect at the parity-invariant Fermi surface, Phys. Rev. B **107**, 125153 (2023).

[32] Shun-Qing Shen, Half quantized Hall effect, Coshare Science **02**, 1 (2024).

[33] P. E. Blöchl, Projector augmented-wave method, Phys. Rev. B **50**, 17953 (1994).

[34] G. Kresse and J. Furthmüller, Efficient iterative schemes for *ab initio* total-energy calculations using a plane-wave basis set, Phys. Rev. B **54**, 11169 (1996).

[35] John P. Perdew, Kieron Burke, and Matthias Ernzerhof, Generalized gradient approximation made simple, Phys. Rev. Lett. **77**, 3865 (1996).

[36] Arash A. Mostofi, Jonathan R. Yates, Young-Su Lee, Ivo Souza, David Vanderbilt, and Nicola Marzari, Wannier90: A tool for obtaining maximally-localised Wannier functions, Comput. Phys. Commun. **178**, 685 (2008).

[37] QuanSheng Wu, ShengNan Zhang, Hai-Feng Song, Matthias Troyer, and Alexey A. Soluyanov, WannierTools: An open-source software package for novel topological materials, Comput. Phys. Commun. **224**, 405 (2018).

[38] Stepan S. Tsirkin, High performance Wannier interpolation of Berry curvature and related quantities with WannierBerri code, npj Comput. Mater. **7**, 33 (2021).

[39] Xuelai Li, Feng Rao, Zhitang Song, Kun Ren, Weili Liu, and Zhimei Sun, Experimental and theoretical study of silicon-doped $Sb_2Te_3$ thin films: Structure and phase stability, Appl. Surf. Sci. **257**, 4566 (2011).

[40] Stefan Grimme, Jens Antony, Stephan Ehrlich, and Helge Krieg, A consistent and accurate ab initio parametrization of density functional dispersion correction (DFT-D) for the 94 elements H-Pu, J. Chem. Phys. **132**, 154104 (2010).

[41] Shuichi Nosé, A unified formulation of the constant temperature molecular dynamics methods, J. Chem. Phys. **81**, 511 (1984).

[42] See the Supplemental Material at http://link.aps.org/, which includes details of calculations with relevant references [33–41], and supplemental figures. .

[43]     M. M. Otrokov, I. I. Klimovskikh, H. Bentmann, D. Estyunin, A. Zeugner, Z. S. Aliev, S. Gaß, A. U. B. Wolter, A. V. Koroleva, A. M. Shikin *et al.*, Prediction and observation of an antiferromagnetic topological insulator, Nature **576**, 416 (2019).

[44]     Yan Gong, Jingwen Guo, Jiaheng Li, Kejing Zhu, Menghan Liao, Xiaozhi Liu, Qinghua Zhang, Lin Gu, Lin Tang, Xiao Feng *et al.*, Experimental realization of an intrinsic magnetic topological insulator, Chinese Phys. Lett. **36**, 076801 (2019).

[45]     M. M. Otrokov, I. P. Rusinov, M. Blanco-Rey, M. Hoffmann, A. Yu. Vyazovskaya, S. V. Eremeev, A. Ernst, P. M. Echenique, A. Arnau, and E. V. Chulkov, Unique thickness-dependent properties of the van der Waals interlayer antiferromagnet $MnBi_2Te_4$ films, Phys. Rev. Lett. **122**, 107202 (2019).

[46]     Jiaheng Li, Yang Li, Shiqiao Du, Zun Wang, Bing-Lin Gu, Shou-Cheng Zhang, Ke He, Wenhui Duan, and Yong Xu, Intrinsic magnetic topological insulators in van der Waals layered $MnBi_2Te_4$-family materials, Sci. Adv. **5**, eaaw5685 (2019).

[47]     Dongqin Zhang, Minji Shi, Tongshuai Zhu, Dingyu Xing, Haijun Zhang, and Jing Wang, Topological axion states in the magnetic insulator $MnBi_2Te_4$ with the quantized magnetoelectric effect, Phys. Rev. Lett. **122**, 206401 (2019).

[48]     Haijun Zhang, Chao-Xing Liu, Xiao-Liang Qi, Xi Dai, Zhong Fang, and Shou-Cheng Zhang, Topological insulators in $Bi_2Se_3$, $Bi_2Te_3$ and $Sb_2Te_3$ with a single Dirac cone on the surface, Nature Phys. **5**, 438 (2009).

[49]     Cong-xin Xia, Juan Du, Xiao-wei Huang, Wen-bo Xiao, Wen-qi Xiong, Tian-xing Wang, Zhong-ming Wei, Yu Jia, Jun-jie Shi, and Jing-bo Li, Two-dimensional n-InSe/p-GeSe(SnS) van derWaals heterojunctions: High carrier mobility and broadband performance, Phys. Rev. B **97**, 115416 (2018).

[50]     Kyungwha Park, J. J. Heremans, V. W. Scarola, and Djordje Minic, Robustness of topologically protected surface states in layering of $Bi_2Te_3$ thin films, Phys. Rev. Lett. **105**, 186801 (2010).

[51]     Minsung Kim, Choong H. Kim, Heung-Sik Kim, and Jisoon Ihm, Topological quantum phase transitions driven by external electric fields in $Sb_2Te_3$ thin films, Proc. Natl. Acad. Sci. USA **109**, 671 (2012).

[52]     Jianwei Sun, Adrienn Ruzsinszky, and John P. Perdew, Strongly constrained and appropriately normed semilocal density functional, Phys. Rev. Lett. **115**, 036402 (2015).

[53]     Guohua Cao, Huijun Liu, Jinghua Liang, Long Cheng, Dengdong Fan, and Zhenyu Zhang, Rhombohedral $Sb_2Se_3$ as an intrinsic topological insulator due to strong van der Waals interlayer coupling, Phys. Rev. B **97**, 075147 (2018).

[54]	Yaohua Liu, Lin-Lin Wang, Qiang Zheng, Zengle Huang, Xiaoping Wang, Miaofang Chi, Yan Wu, Bryan C. Chakoumakos, Michael A. McGuire, Brian C. Sales *et al.*, Site Mixing for Engineering Magnetic Topological Insulators, Phys. Rev. X **11**, 021033 (2021).

[55]	Xi Wu, Chao Ruan, Peizhe Tang, Feiyu Kang, Wenhui Duan, and Jia Li, Irremovable Mn-Bi Site Mixing in MnBi$_2$Te$_4$, Nano Letters **23**, 5048 (2023).

[56]	Yugui Yao, Leonard Kleinman, A. H. MacDonald, Jairo Sinova, T. Jungwirth, Ding-sheng Wang, Enge Wang, and Qian Niu, First principles calculation of anomalous Hall conductivity in ferromagnetic bcc Fe, Phys. Rev. Lett. **92**, 037204 (2004).

[57]	Ruie Lu, Hongyi Sun, Shiv Kumar, Yuan Wang, Mingqiang Gu, Meng Zeng, Yu-Jie Hao, Jiayu Li, Jifeng Shao, Xiao-Ming Ma *et al.*, Half-Magnetic Topological Insulator with Magnetization-Induced Dirac Gap at a Selected Surface, Phys. Rev. X **11**, 011039 (2021).

[58]	Dmitri K. Efetov and Philip Kim, Controlling electron-phonon interactions in graphene at ultrahigh carrier densities, Phys. Rev. Lett. **105**, 256805 (2010).

[59]	J. T. Ye, Y. J. Zhang, R. Akashi, M. S. Bahramy, R. Arita, and Y. Iwasa, Superconducting dome in a gate-tuned band insulator, Science **338**, 1193 (2012).

[60]	Paul V. Nguyen, Natalie C. Teutsch, Nathan P. Wilson, Joshua Kahn, Xue Xia, Abigail J. Graham, Viktor Kandyba, Alessio Giampietri, Alexei Barinov, Gabriel C. Constantinescu *et al.*, Visualizing electrostatic gating effects in two-dimensional heterostructures, Nature **572**, 220 (2019).

[61]	Feng Xue, Zhe Wang, Yusheng Hou, Lei Gu, and Ruqian Wu, Control of magnetic properties of MnBi$_2$Te$_4$ using a van der Waals ferroelectric III$_2$-VI$_3$ film and biaxial strain, Phys. Rev. B **101**, 184426 (2020).

[62]	Jing-Yang You, Xue-Juan Dong, Bo Gu, and Gang Su, Electric field induced topological phase transition and large enhancements of spin-orbit coupling and Curie temperature in two-dimensional ferromagnetic semiconductors, Phys. Rev. B **103**, 104403 (2021).